\documentclass[a4paper,12pt]{article}
\usepackage{epsfig}
\usepackage[american]{babel}
\usepackage{amssymb}

\begin{document}
\title{On possibility of realization of the phenomena of complex analytic
dynamics in physical systems. Novel mechanism of the
synchronization loss in coupled period-doubling systems}
\author{Olga~B.~Isaeva$^*$, Sergey~P.~Kuznetsov}
\date{}
\maketitle\begin{center} \emph{ Institute of Radio-Engineering and
Electronics of RAS, Saratov Branch, \\ Zelenaya 38, Saratov,
410019, Russia\\
$^*$E-mail:IsaevaOB@info.sgu.ru}\end{center}

\maketitle

\begin{abstract}
The possibility of realization of the phenomena of complex
analytic dynamics for the realistic physical models are
investigated. Observation of the Mandelbrot and Julia sets in the
parameter and phase spaces both for the discrete maps and
non-autonomous continuous systems is carried out. For these
purposes, the method, based on consideration of coupled systems,
demonstrating period-doubling cascade is suggested. Novel
mechanism of synchronization loss in coupled systems corresponded
to the dynamical behavior intrinsic to the complex analytic maps
is offered.
\end{abstract}

\section{Introduction}
It is known~\cite{Peitgen,Devaney}, that complex analytic dynamics
(CAD), studying behavior of complex maps, includes a lot of
interesting phenomena, for example, presence of fractal Mandelbrot
and Julia sets in the parameter and phase spaces.

Let us start with a quadratic logistic map
\begin{equation}\label{eq1}
{z}' \to \lambda - z^{2},
\end{equation}
where $\lambda$ is a complex parameter, and $z$ is a complex
variable. By definition, Mandelbrot set (fig.~1$ð$) is a set of
points on a plane of complex parameter $\lambda$, for which the
orbit of an extremum  $z=0$ of the map~(\ref{eq1}) during
iteration procedure does not escape to infinity. The Mandelbrot
set contains the so-called "Mandelbrot cactus" (designed on figure
by gray color); this is a set of points in the parameter plane,
for which the trajectory starting of the extremum of the map
converges to a periodic attractor.

%==================================================%
\begin{figure}
\centerline{\epsfig{file=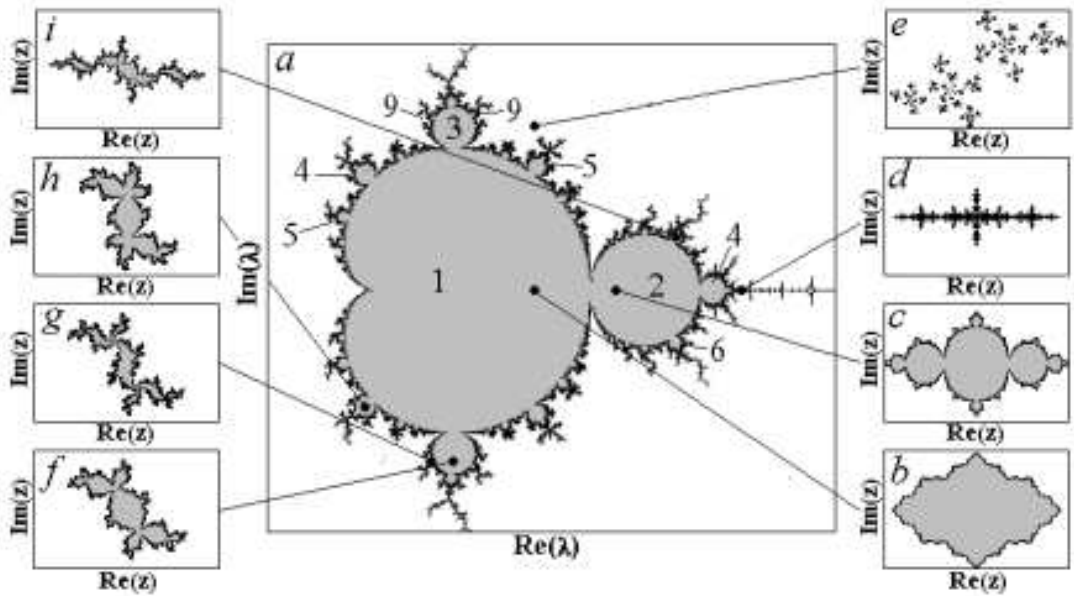,width=0.8\textwidth}}

\caption{Mandelbrot set (a) and Julia sets for the quadratic
complex map with different values of parameter: $\lambda=0.5$ (b),
$\lambda=0.8$ (c), $\lambda=1.42$ (d), $\lambda=0.5+0.7i$ (e),
$\lambda=0.123-0.745i$ (f), $\lambda=0.0315-0.7908i$ (g),
$\lambda=-0.282+0.530i$ (h), $\lambda=1.16+0.25i$ (i). The gray
color designates the regions corresponding to existence of
periodic dynamics (periods are marked by respective numbers); the
black color designates points, at which the restricted in a phase
space chaotic dynamics is implemented; the white color means the
escaping of trajectories to infinity.} \label{fig1}
\end{figure}
%==================================================%

"Mandelbrot cactus" consists of a big cardioid, corresponding to
the existence of an attracting fixed point, and an infinite number
of "leaves", corresponding to existence of attracting cycles of
different periods. For example, the leaves of the doubled periods
are placed along a real axis. The sequences of other period
m-tupling (period-multiplication) bifurcations also can be found.
In particular, the accumulation points for period-tripling and
period-quadrupling bifurcation cascades have been investigated in
the work of Golberg, Sinai and Khanin~\cite{Golberg}. In the works
of Cvitanovi\'{c} and Mirheim~\cite{Cvitanovic1,Cvitanovic2} the
universal properties of many other bifurcation cascades were
studied.

A bifurcation, which is responsible for originating a "leave",
corresponds to a stability loss of a "parent" cycle, characterized
by a complex multiplier with unit modulus and rational argument
(in relation to $2\pi$). If the argument of a multiplier at the
stability loss is irrational, then the domains with fractal
boundaries, filled by invariant curves arise in the phase plane,
the so-called Siegel disks~\cite{Widom,Manton,MacKay}.

Fractal pattern close to the "Mandelbrot cactus" and denoted by
black color in Fig.~1, corresponds to existence of the chaotic
dynamical regimes in the phase space.

In figures~1($b$-$i$) the Julia sets for different values of
complex parameter $\lambda$ are shown. The Julia set is a border
between basins of attraction to infinity (white color) and to a
periodic motion (gray color) in a plane of complex variable $z$.
One can distinguish the following types of Julia sets:

\begin{itemize}
  \item for values $\lambda$, belonging to the "Mandelbrot cactus", the Julia
set is connected and enclose an interior basin
(figs.~1($b$,$c$,$f$-$i$);
  \item for values $\lambda$, at which chaotic dynamics exists, the Julia
set is also connected, but has no inner region (fig.~1$d$);
  \item for values $\lambda$, outside the Mandelbrot set, the Julia set is
disconnected (fig.~1$f$).
\end{itemize}

It is obvious, that 1D complex map can be represented equivalently
by a 2D real map (for this purpose it is necessary only to
separate real and imaginary parts of the equation). However, the
mentioned phenomena of CAD are intrinsic only to a very special
class of the real 2D maps, namely for the analytic maps, obeying
the Cauchy-Riemann conditions. Violation of the analyticity leads
to drastic changes of the dynamics of the
map~\cite{rcd,Peinke,Klein,Peckham1,Peckham2}. Thereby, a
following problem arises: Is it possible to specify actual
physical systems demonstrating phenomena of CAD? Recently, this
problem attracts great attention. The physical applications of
complex dynamics for such problems, as the renormalization group
approach in the theory of phase transitions and the theory of a
percolation were discussed~\cite{Hu,percol,Abdusalam,npcs}. In the
paper of Beck~\cite{Beck} a theoretical possibility of the
construction of the physical system, in which the Mandelbrot set
would arise, was considered. The suggested approach is based on
analysis of a motion of a charged particle in a double-peak
potential with non-linear damping. The particle is driven by
magnetic field, depending on time and on the particle velocity,
and effected by external shot pulses, time intervals between which
also depend on the particle velocity.

In present work, we suggest a simpler and universal approach of
constructing models manifesting the Mandelbrot set and other
phenomena of CAD, which may be designed as realistic physical
systems. This method allow us to carry out a physical
experiment and present the first observation of the Mandelbrot
set~\cite{Isaeva}. The special structure of the Fourier spectrum
of signal generated by experimental system at the period-tripling
accumulation point is presented in~\cite{Isaeva2}.

Our method is based on using of specially symmetrically coupled
identical systems, demonstrating transition to chaos through
period-doublings. As it is known, such behavior is peculiar for a
very wide class of nonlinear dissipative systems of various
physical nature. The special kind of offered coupling provides a
special symmetry in the multi-dimensional system, which is
necessary for implementation of the analyticity conditions. It is
a simple problem to construct the system with such coupling in
comparison to the system, suggested by Beck.

In section 2 the procedure of construction of the coupled systems
demonstrating phenomena of CAD at the example of the discrete
logistic maps is considered. In sections 3 and 4 we apply
developed method to the various systems of Feigenbaum universality
class, namely to the H\'{e}non map and to the nonlinear
non-autonomous oscillator. The relation of the CAD phenomena to
the phenomena of generalized partial synchronization is discussed
in section 5.

\section{From complex quadratic map to coupled logistic maps}
Let us start with the notion that one-dimensional complex
quadratic map is equivalent to the  system of two real coupled
quadratic maps with a special type of coupling.

Separation of real and imaginary parts in the complex equation
(\ref{eq1}) yields
\begin{equation}
\label{eq2} z'_{re} \to \lambda _{re} - z_{re}^{2} + z_{im}^{2},
\quad z'_{im} \to \lambda _{im} - 2z_{re} z_{im}.
\end{equation}
Next, we introduce the following designations
\begin{equation}
\label{eq3}
\begin{array}{c}
 x_1=z_{re}+\beta z_{im}, \quad x_2=z_{re}-\beta z_{im}, \\
 \lambda_1=\lambda_{re}+\beta \lambda_{im}, \quad \lambda_2=\lambda_{re}-\beta \lambda_{im}.
 \end{array}
\end{equation}
As a result we obtain a system of two coupled quadratic maps
\begin{equation}
\label{eq4}
\begin{array}{c}
x'_{1} \to \lambda_1-x_{1}^{2}+\varepsilon (x_2-x_1)^2, \\ x'_{2}
\to \lambda_2-x_{2}^{2}+\varepsilon (x_1-x_2)^2, \\
\end{array}
\end{equation}
where $\varepsilon=(1+\beta^2)/4\beta^2$ is the parameter of
coupling. Note a special type of coupling in these equations. It
can be interpreted as an identical simultaneous shift of control
parameters in both partial systems, proportional to the squared
difference of dynamic variables at each step of discrete time.

It is worth nothing that the coefficient
$\varepsilon=(1+\beta^{2})/4\beta^{2}$ for any $\beta$ is larger
than $1/4$. Nevertheless, formally we can investigate
system~(\ref{eq4}) with any $\varepsilon$.

In fig.~2 we present the charts of the parameter plane
$(\lambda_1, \lambda_2)$ for the coupled maps~(\ref{eq4}) at
several values of parameter $\varepsilon$. One can see the usual
Mandelbrot set, rotated by $45^{\circ}$, takes place at
$\varepsilon=0.5$. For $0.25<\varepsilon<+\infty$ we have a
distorted Mandelbrot set on the parameter plane. The cactus leaves
of this Mandelbrot set correspond to existence of periodic motion
of different periods. At $\varepsilon=0.25$, the set on the
parameter plane, for which the point starting from the origin does
not escape to infinity, looks like a set of strips, where the
period doubling cycles occur. At $\varepsilon<0.25$ it transforms
to a rhombus-like structure. At a particular $\varepsilon=0$
(uncoupled logistic maps) it is a square.

The generalization of coupling to the case of
$-\infty<\varepsilon<+\infty$ corresponds to the original
map~(\ref{eq1}), variable and parameter of which are so-called
two-component
numbers~\cite{Lavrentjev,Senn,Fjelstad,Ronveaux,Majernic,Band}.
This is a special algebraic system, which elements are defined as
follows
\begin{equation}\label{twocom}
  z=x+iy, \quad i^{2}=a+ib, \quad \mathrm{where} \quad a,b \in \bf R.
\end{equation}

%==================================================%
\begin{figure}
\centerline{\includegraphics[width=0.8\textwidth,keepaspectratio]{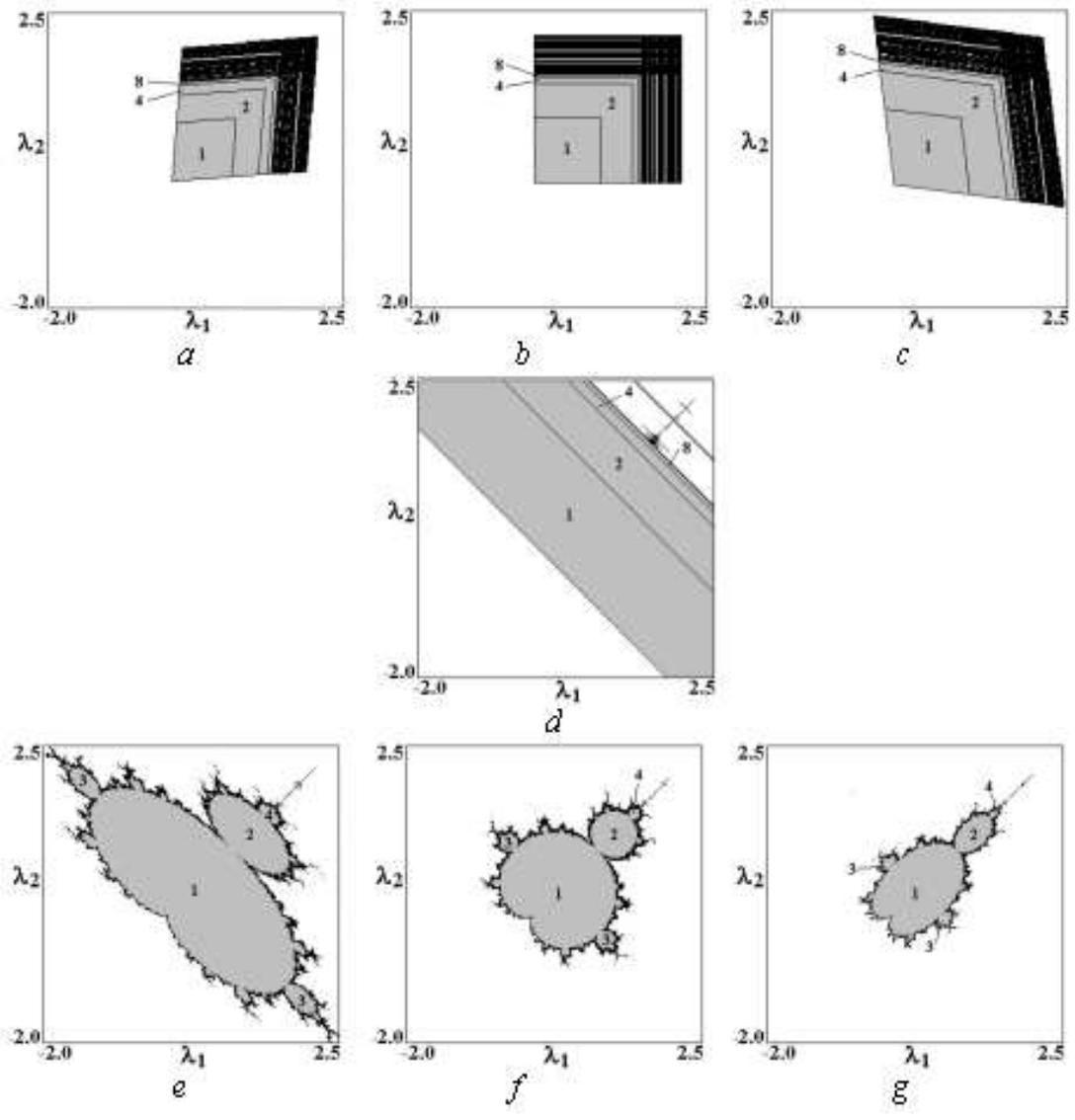}}
\caption{The charts of the parameter plane $(\lambda_1,
\lambda_2)$ for the coupled logistic maps~(\ref{eq4}) with
different values of parameter of coupling: $\varepsilon=-0.1$ (a),
$\varepsilon=0.0$ (b), $\varepsilon=0.1$ (c), $\varepsilon=0.25$
(d), $\varepsilon=0.3$ (e), $\varepsilon=0.5$ (f),
$\varepsilon=1.0$ (g). The figures (a-c) correspond to hyperbolic
numbers, figure (d) -- to parabolic, and figures (e-g) -- to
elliptic numbers.}
\end{figure}
%==================================================%

According to~\cite{Lavrentjev}, there are three special cases:
$i^2=-1$ -- the usual complex numbers, $i^2=+1$ -- the so-called
perplex numbers, $i^2=0$ -- the dual numbers. All other algebraic
number systems are isomorphic to complex, perplex or dual numbers
depending on, whether the value of $(a+b^2)/4b^2$ is positive,
negative or zero, and are known as elliptic, hyperbolic or
parabolic number system, respectively. In terms of parameter
$\varepsilon$ these conditions look as follows:

\begin{description}
\item[1)] the case $\varepsilon>0.25$ corresponds to
elliptic numbers isomorphic to complex numbers, implemented at
$\varepsilon=0.5$;
\item[2)] the case $\varepsilon<0.25$ corresponds to hyperbolic numbers isomorphic to perplex numbers, implemented at
$\varepsilon=0$;
\item[3)] the case $\varepsilon=0.25$ corresponds to parabolic or dual
numbers.
\end{description}

Thus, the existence of three topologically different structures on
the plane of parameters $(\lambda _1,\lambda_2)$, namely, fractal
structure, similar to Mandelbrot set, rhombus-like structure and
system of strips, is explained by existence of three different
algebraic systems of numbers.

\section{Coupled H\'{e}non maps}
The more realistic model, than the logistic map, is the
H\'{e}non map~\cite{Henon1,Henon2}
\begin{equation}
\label{eq5}
\begin{array}{c}
x' \to \lambda-x^2-d \cdot y, \\ y' \to x,
\end{array}
\end{equation}
which is two-dimensional reversible map and, therefore, can be
realized as the Poincare cross-section of some continuous system
with 3-D phase space -- the minimal dimensionality ensuring an
possibility of nontrivial dynamics and chaos.

In present section we show, that it is possible to observe the
phenomena of CAD in the system of two coupled H\'{e}non maps.
(In~\cite{Isaeva_h} it is shown, that phenomena of CAD such as period
multiplication cascades survives in complex H\'{e}non map)
For this purpose we carry out the following discourse, which
starting point is the complexification of the individual map
(\ref{eq5}).

Let us consider dynamical variables $x$ and $y$ and driving
parameter $\lambda$ responsible to transition to chaos through
period-doublings as complex, that is $x=x_{re}+ix_{im}$,
$y=y_{re}+iy_{im}$, $\lambda=\lambda_{re}+i\lambda_{im}$.
Parameter $d$ we consider as real. Separating real and imaginary
parts, we obtain:
\begin{equation} \label{eq6}
\begin{array}{c}
 {x}'_{re} \to \lambda_{re}-x_{re}^{2}+x_{im}^{2}-d \cdot y_{re},
\\
 {x}'_{im} \to \lambda_{im}-2x_{re} x_{im}-d \cdot y_{im}, \\
 {y}'_{re} \to x_{re} , \\
 {y}'_{im} \to x_{im}. \\
 \end{array}
\end{equation}
Let us enter new variables and parameters
\begin{equation}
\label{eq7}
\begin{array}{c}
x_{1,2}=x_{re} \pm \beta x_{im}, \\ y_{1,2}=y_{re} \pm \beta
y_{im}, \\ \lambda_{1,2}=\lambda_{re} \pm \beta \lambda_{im}, \\
\varepsilon=(1+\beta^{2})/4\beta^{2}.
 \end{array}
\end{equation}
Then, the complexified H\'{e}non map, represented in the form
of coupled real maps looks as follows
\begin{equation}
\label{eq8}
\begin{array}{c}
{x}'_1 \to \lambda_1-x_{1}^{2}-d \cdot y_{1}+\varepsilon
(x_2-x_1)^2, \\ {y}'_1 \to x_1, \\ {x}'_2 \to \lambda
_2-x_{2}^{2}-d \cdot y_2+\varepsilon (x_1-x_2)^2, \\ {y}'_2 \to
x_2.
\end{array}
\end{equation}

\begin{figure}
\centerline{\includegraphics[width=0.8\textwidth,keepaspectratio]{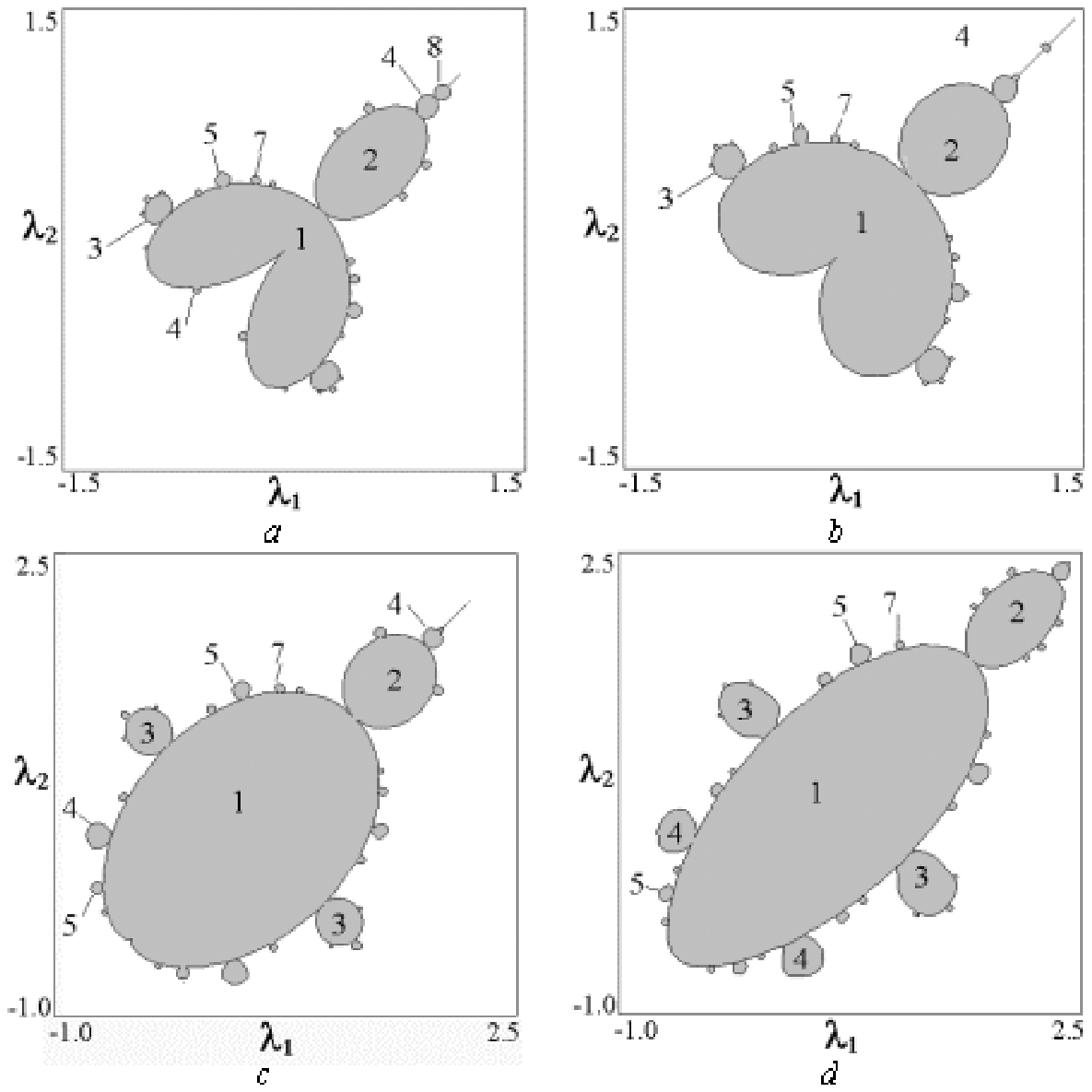}}
\caption{The Mandelbrot set for coupled H\'{e}non mappings
(\ref{eq8}) with the following values of parameter $d$: $-0.5$
(a); $-0.3$ (b); $0.3$ (c); $0.5$ (d). Parameter of coupling
$\varepsilon=0.5$.}

\centerline{\includegraphics[width=0.9\textwidth,keepaspectratio]{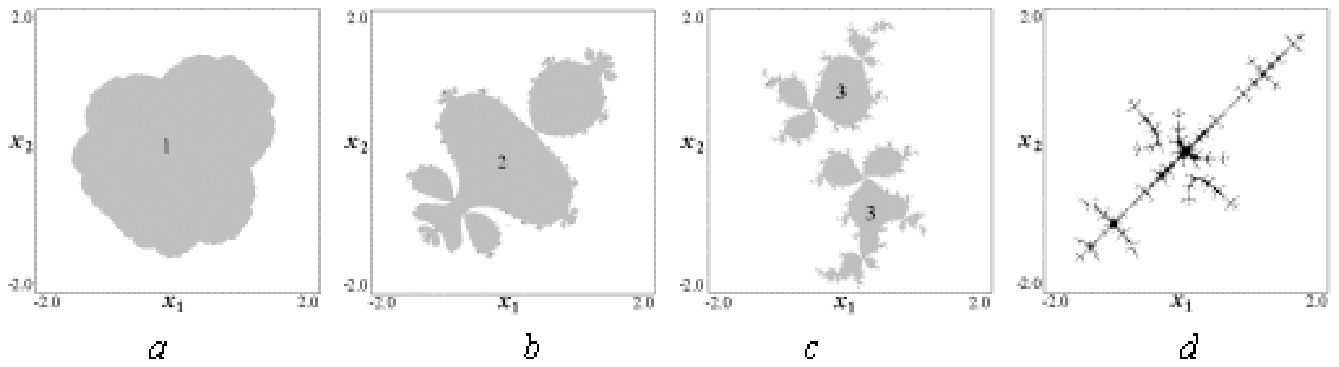}}
\caption{The Julia sets for coupled H\'{e}non mappings with
$d=-0.3$ for the following values of parameters $\lambda_1$ and
$\lambda_2$: $\lambda_1=\lambda_2=0.1$ (a);
$\lambda_1=\lambda_2=0.5$ (b); $\lambda_1=-0.8, \lambda_2=0.5$
(c); $\lambda_1=\lambda_2=1.05$ (d).}
\end{figure}

At fig.~3 the charts of the plane of parameters
$(\lambda_1,\lambda_2)$ for various values of parameter $d$ are
shown. As one can see, the parameter plane contains the domains
similar by the shape to the Mandelbrot set, distorting with
increasing of $|d|$. At the next figure the Julia sets for
system~(\ref{eq8}) are represented. They also have apparent
similarity to usual Julia sets. The basins of attraction of the
fixed point~(fig.~4a), cycles of period 2~(fig.~4b) and
3~(fig.~4c) corresponding to the filled-in Julia sets, and basin
of attraction of chaotic attractor~(fig.~4d) corresponding to the
connected dendrit-like Julia set are shown. All figures in present
and next sections are corresponded to the value of coupling
parameter $\varepsilon=0.5$ -- the value, which is equivalent to
usual complex numbers and usual Mandelbrot set.

\section{Coupled nonlinear oscillators}
One of the most universal models, suitable for description of a
number of real physical systems, is the nonlinear oscillator. Let
us consider a quadratic oscillator with damping and harmonic
driving force
\begin{equation}
\label{eq9} \ddot {x}+\gamma \dot {x}+\lambda+x^{2}=F\cos \omega
t,
\end{equation}
where $x$ -- dynamical variable,$\lambda$ -- parameter of
nonlinearity, $\gamma$ -- parameter of damping, $F$ and $\omega$
-- amplitude and frequency of external driving force. It is known,
that such systems demonstrate the opportunity of transition to
chaos through a cascade of period-doubling bifurcations, for
example by changing of parameter $\lambda$ with fixed $F$,
$\omega$ and $\gamma$~\cite{Moon,Potapova} (see~fig.~5a).

Let us construct system of two coupled oscillators demonstrating
phenomena of CAD, operating the same scheme as in the previous
sections. At first, we complexify the equation of a quadratic
oscillator in such way, that driving parameter $\lambda$ and
variable $x$ are complex, and then, we make following designations
\begin{equation}
\label{eq10}
\begin{array}{c}
x_{1,2}=x_{re} \pm \beta x_{im}, \\

\lambda_{1,2}=\lambda_{re} \pm \beta \lambda_{im}, \\

\varepsilon=(1+\beta^{2}) / 4 \beta^{2}.
\end{array}
\end{equation}
As a result we obtain the system of coupled oscillators
\begin{equation}
\label{eq11}
\begin{array}{c}
\ddot {x}_1+\gamma \dot {x}_1+\lambda_{1}+x_{1}^{2}-\varepsilon
(x_2-x_1)^2=F \cos \omega t, \\ \ddot {x}_2+\gamma \dot
{x}_2+\lambda_{2}+x_{2}^{2}-\varepsilon (x_1-x_2)^2=F \cos \omega
t. \\
\end{array}
\end{equation}

At fig.~5b the chart of a plane of parameters
$(\lambda_1,\lambda_2)$ is exhibited. Although, the represented
picture differ by the shape from the Mandelbrot set for coupled
logistic maps, but it have its basic properties, such as presence
of leaves of every possible periods and their self-similar
organization. The difference is determined by the existence of the
region of bistability for the quadratic oscillator~(see~fig.~5a).

Analogy between basins of periodic attractors for coupled
oscillators represented at fig.~6 and usual Julia sets is also
obvious. With $\lambda_1=\lambda_2=-0.8$ the basin of one
attracting point is realized~(figs.~6a). Then, it splits up to two
basins of different fixed points (fig.~6b). With
$\lambda_1=\lambda_2=-0.575$ (fig.~6c) one can see, that one of
the basins destroys and with $\lambda_1=\lambda_2=-0.5$ there is
only one attracting fixed point on the phase plane (fig.~6d).
Figs.~6e-f demonstrate the basins of attracting cycles of period 2
and 3.  It is necessary to note, that for non-autonomous coupled
oscillators there can be a number of basins of attraction of
various periodic and chaotic motion on the phase plane
$(x_1,x_2)$. Thus, only one of them, posed at a central part of
figures is associated with Julia set corresponded to the
Mandelbrot set of figure~5b.

\begin{figure}
\centerline{\includegraphics[width=0.4\textwidth,keepaspectratio]{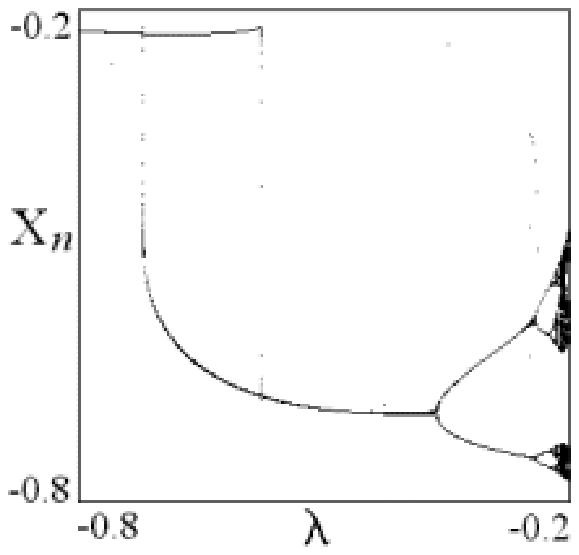}\hspace{2cm}
\includegraphics[width=0.4\textwidth,keepaspectratio]{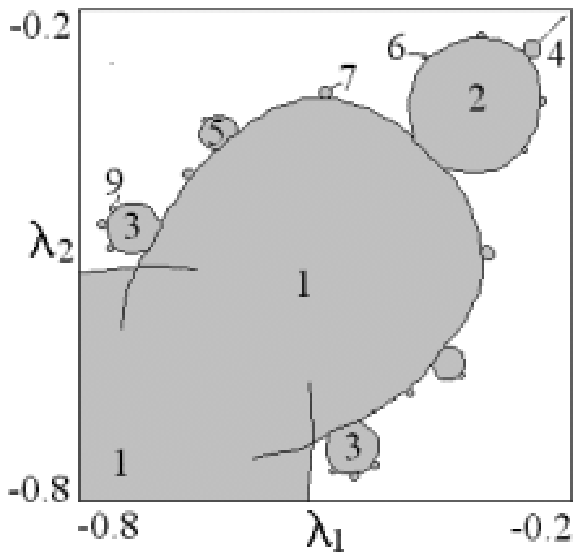}}
\centerline{(a)\hspace{9cm}(b)}

\caption{(a): Bifurcation tree for the quadratic oscillator
(\ref{eq9}) at the plane $(X_n, \lambda)$, where $X_n$ are the
values of dynamical variable $x$ at the Poincare cross-section
$t=2\pi n$ ($n=1,2,...$). Cascade of period-doubling bifurcations
and region of multi-stability is visible. (b): The charts of
dynamical regimes on a plane of parameters $(\lambda_1,
\lambda_2)$ for system of coupled oscillators (\ref{eq11}) with
$\gamma=0.2$, $F=0.23$, $\omega=1$, $\varepsilon=0.5$.}

\centerline{\includegraphics[width=0.9\textwidth,keepaspectratio]{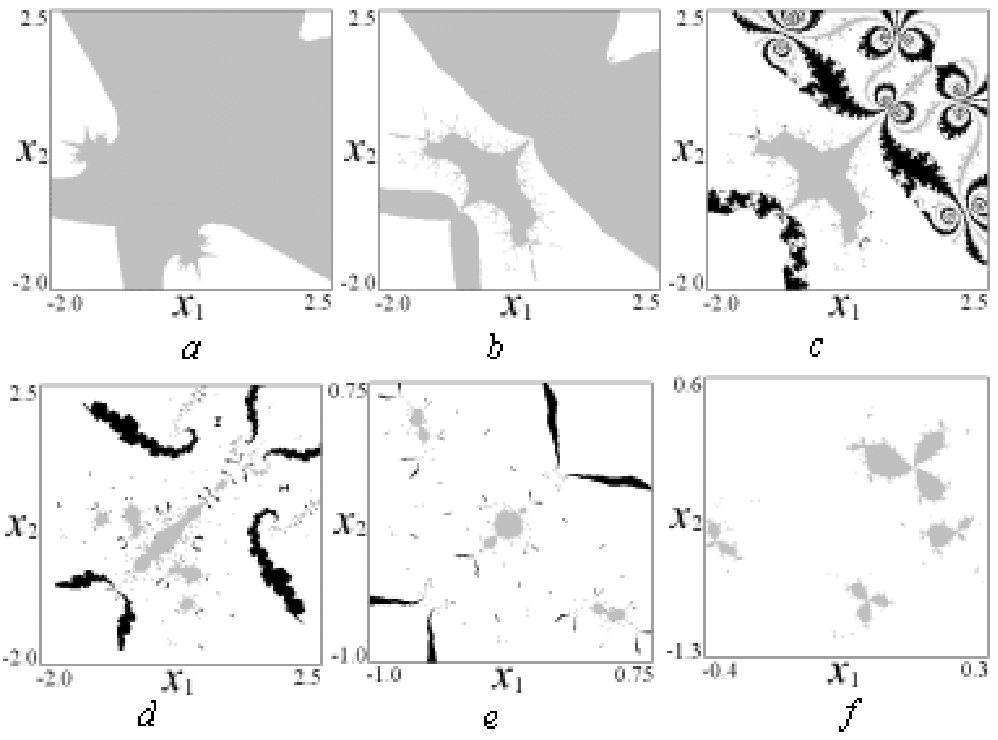}}
\caption{The charts of the phase space cross-section $(x_1,x_2)$
for coupled quadratic oscillators with following values of
parameters: $\lambda_1=\lambda_2=-0.8$ (a),
$\lambda_1=\lambda_2=-0.6$ (b), $\lambda_1=\lambda_2=-0.575$ (c),
$\lambda_1=\lambda_2=-0.5$ (d), $\lambda_1=\lambda_2=-0.3$ (e),
$\lambda_1=-0.74, \lambda_2=-0.45$ (f). The basins of attraction
of the fixed point (figs. a-d), and cycles of period 2 (fig. e)
and 3 (fig. f) at the Poincare cross-section are represented.}
\end{figure}

\begin{figure}
\centerline{\includegraphics[width=0.7\textwidth,keepaspectratio]{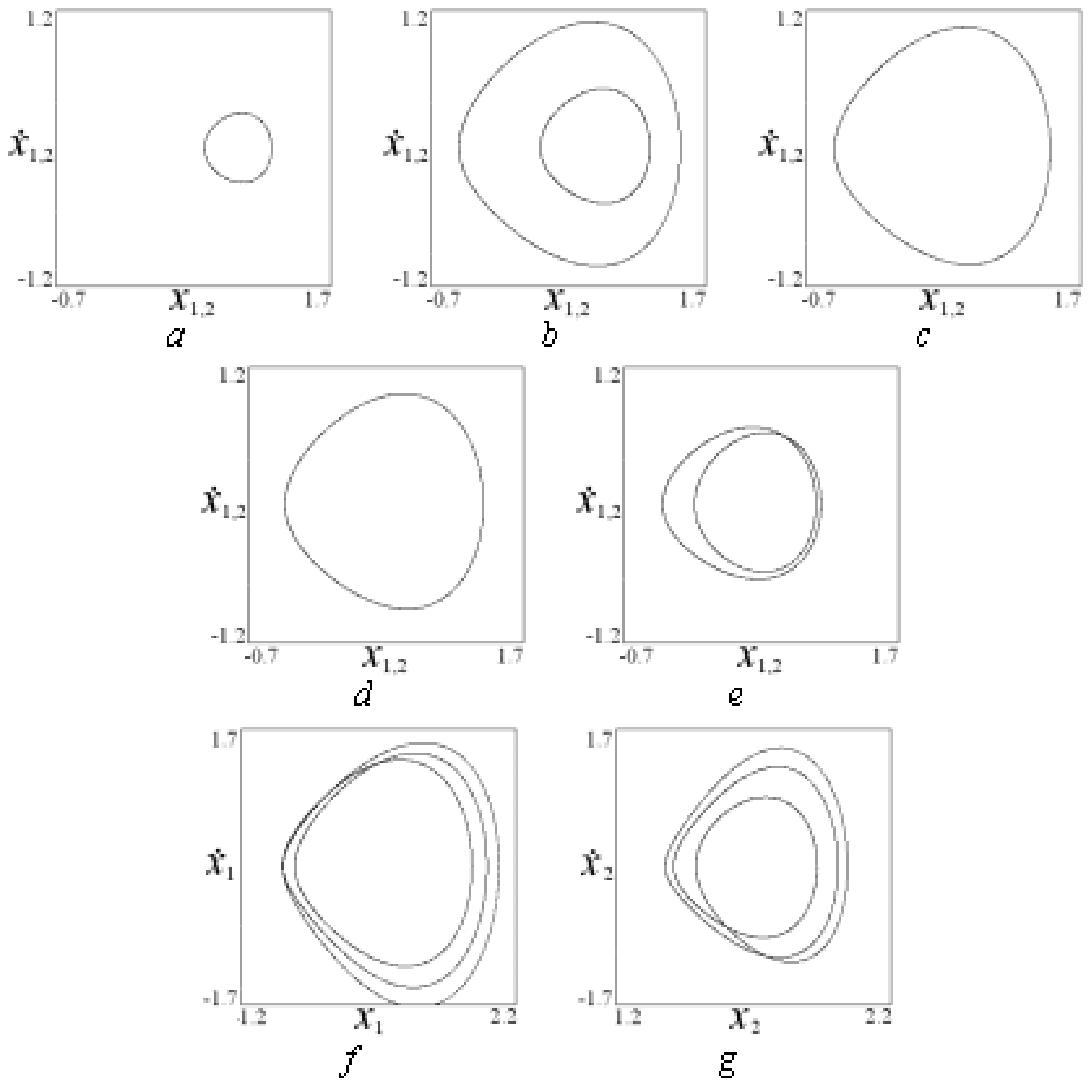}}
\caption{The charts of the phase plane $(x_1,\dot x_1)$ and
$(x_2,\dot x_2)$ for coupled quadratic oscillators with following
values of parameters: $\lambda_1=\lambda_2=-0.8$ (a),
$\lambda_1=\lambda_2=-0.6$ (b), $\lambda_1=\lambda_2=-0.575$ (c),
$\lambda_1=\lambda_2=-0.5$ (d), $\lambda_1=\lambda_2=-0.3$ (e),
$\lambda_1=-0.74, \lambda_2=-0.45$ (f).}

\centerline{\includegraphics[width=0.7\textwidth,keepaspectratio]{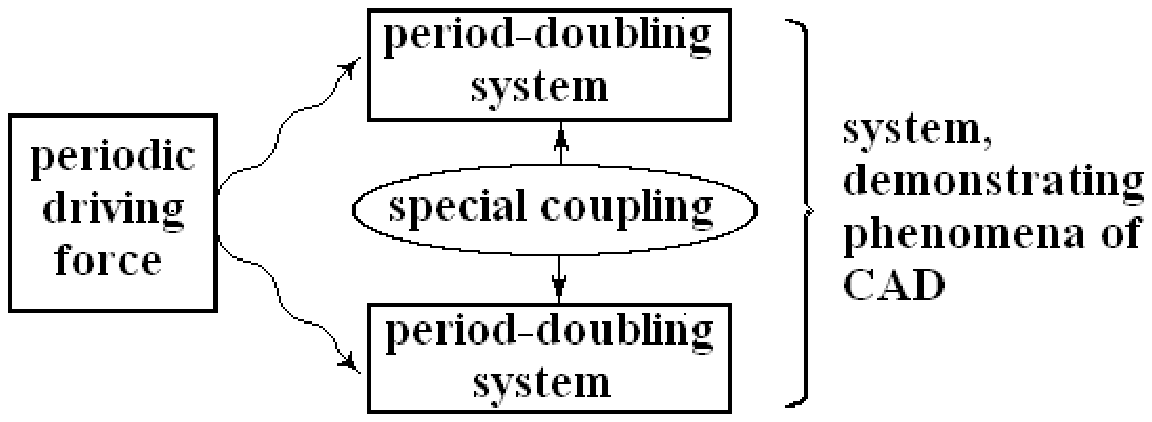}}
\caption{Scheme of construction of the physical system with CAD
phenomena.}
\end{figure}

Figure~7 represents the phase planes $(x_1,\dot x_1)$, $(x_2,\dot
x_2)$ with the same values of parameters as for the fig.~6.

Thus, the strategy of construction of the flow system manifesting
CAD phenomena is follow (see scheme at fig.~8): One should take
two identical elements (describable by the continuous differential
equations), make special coupling between them and synchronize
them by common periodic external driving force.

\section{Phenomena of CAD and mechanisms of synchronization
loss of coupled systems} The investigation of phenomenon of
synchronization loss in coupled systems demonstrating transition
to chaos through period-doubling bifurcations is one of the mostly
interesting problems of nonlinear dynamics and has a great
fundamental and applied importance for different fields of science
and technique. It is necessary to note, that the set of domains of
periodic dynamics corresponded to leaves of Mandelbrot cactus, can
be considered as the region of generalized partial
synchronization. By generalized partial synchronization we mean
the dynamical state of a system, in which the value
$(x_{1}^{2}+x_{2}^{2})^{1/2}$ is bounded and varies periodically.
It means, that the trajectories of one subsystem do not escape far
from trajectories of other subsystem. At fig.~9 the charts of
phase planes for coupled logistic maps with various values of
$\lambda_1$ and $\lambda_2$ are represented. At these pictures,
the bold dots indicate the periodic attractors, which basins of
attraction are marked by the gray color. It is easy to see, that
in the case $\lambda_1=\lambda_2$ there is full synchronization in
the system (attractor takes place at the diagonal line). In the
case, when the parameters of subsystems $\lambda_1$ and
$\lambda_2$ are not equal, but belongs to the "Mandelbrot cactus",
the generalized synchronization of subsystems exists (periodic
attractor takes place at the neighborhood of a diagonal line).

Let us demonstrate the idea of generalized partial synchronization
also with an example of phase portraits of coupled oscillators.
One can see, that with identical values of parameters $\lambda_1$
and $\lambda_2$ (see fig.~8a-e) the values of variables in
subsystems are coincided, that corresponds to full
synchronization. In a case $\lambda_1 \ne \lambda_2$ (fig.~8f,g),
when the phenomena of CAD can be realized, the values of dynamical
variables of subsystems do not coincide, that corresponds to
realization of generalized partial synchronization. Thus, the
phenomena of CAD such as period-tripling bifurcations can be
implemented, when the point $(\lambda_1,\lambda_2)$ belongs to
region of generalized partial synchronization.
\begin{figure}
\centerline{\includegraphics[width=0.9\textwidth,keepaspectratio]{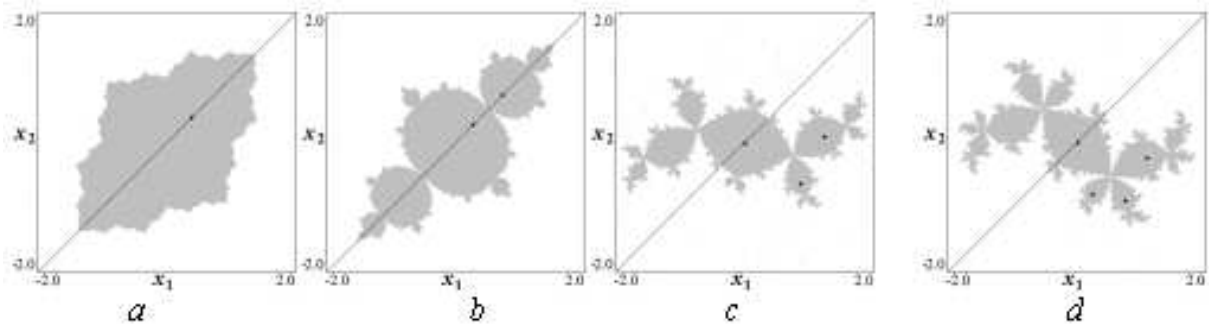}}
\caption{Charts of the phase plane $(x_1, x_2)$ for the coupled
logistic mappings with $\varepsilon=0.5$ for the following values
of parameters $\lambda_1$ and $\lambda_2$:
$\lambda_1=\lambda_2=0.5$ (a); $\lambda_1=\lambda_2=0.8$ (b);
$\lambda_1=0.868, \lambda_2=0.622$ (c); $\lambda_1=0.248,
\lambda_2=0.812$ (d). The dots designate the attractors, which
basins are marked by the gray color. The figures for a fixed point
(a) and cycles of period 2 (b), 3 (c) and 4 (d) are represented.}
\end{figure}

\section{Conclusion}
In present work we have offered a novel universal method for
obtaining of the phenomena of CAD in the realistic models of
physical systems. It is shown, that complexified system can be
represented as two coupled real systems. Such representation is
useful for some reasons. At first, often it simplify equations and
construction of a real physical system. Secondly, with entering of
special coupling to the system of two identical devices of any
nature demonstrating the period-doublings cascade, one may expect
the whole system to demonstrate phenomena of complex analytic
dynamics.

The coupled logistic maps, coupled invertible H\'{e}non maps and
coupled nonlinear oscillators with periodic driving have been
considered. Thus, we have studied both discrete maps, and
non-autonomous system with continuous time.

It has been shown, that realization of the Mandelbrot set in the
parameter space of the systems with discreet time and
non-autonomous periodically driven systems is easy problem enough.
It is necessary to consider two elements demonstrating
period-doublings with coupling, arising from the complexification
of the variables and parameter responsible for the period
doublings in the original system. The special interest is
attracted by the problem of realization of phenomena of CAD in the
autonomous continuous systems.

Besides it is necessary to mark connection of the phenomena of CAD
with a problem of synchronization. It is shown, that Mandelbrot set
corresponds to the domain of generalized partial synchronization of
coupled systems. Thus, the possibility of new nontrivial
features of dynamics corresponded to synchronization of
complex systems has been found.

\section*{Acknowledgements}
This work is supported by RFBR (projects No 03-02-16074 and No 04-02-04011)
and CRDF (REC-006).

\end{document}